%% file: main.tex
\documentclass[10pt,twocolumn,letterpaper]{article}

\usepackage[pagenumbers]{cvpr} 

\input{preamble}

\definecolor{cvprblue}{rgb}{0.21,0.49,0.74}
\usepackage[pagebackref,breaklinks,colorlinks,citecolor=cvprblue]{hyperref}

\def\paperID{176} 
\def\confName{3DV\xspace}
\def\confYear{2026\xspace}

\newcommand{\method}{\textsc{3D-Generalist}\xspace}
\newcommand{\modulea}{Panoramic Environment Generation\xspace}
\newcommand{\moduleb}{Scene-Level Policy\xspace}
\newcommand{\modulec}{Asset-Level Policy\xspace}

\title{3D-GENERALIST: Vision-Language-Action Models for Crafting 3D Worlds}

\author{
Fan-Yun Sun,\hspace{.5em} Shengguang Wu,\hspace{.5em} Christian Jacobsen,\hspace{.5em} Thomas Yim,\hspace{.5em} Haoming Zou,\\
Alex Zook,\hspace{.5em} Shangru Li,\hspace{.5em} Yu-Hsin Chou,\hspace{.5em} Ethem Can,\hspace{.5em} Xunlei Wu,\hspace{.5em} Clemens Eppner, \\
Valts Blukis,\hspace{.5em} Jonathan Tremblay,\hspace{.5em} Jiajun Wu,\hspace{.5em} Stan Birchfield\textsuperscript{\dag},\hspace{.5em} Nick Haber\textsuperscript{\dag}\\\\
$^1$Stanford University, $^2$Nvidia Research \\
{\small \textcolor{blue!70!black}{\tt \href{https://ai.stanford.edu/~sunfanyun/3d-generalist/}{https://ai.stanford.edu/\textasciitilde sunfanyun/3d-generalist/}}} \vspace{-0.2cm}
}

\begin{document}
\twocolumn[{%
\maketitle
\thispagestyle{empty}
\begin{figure}[H]
\hsize=\textwidth
    \centering
   \includegraphics[width=.9\textwidth]{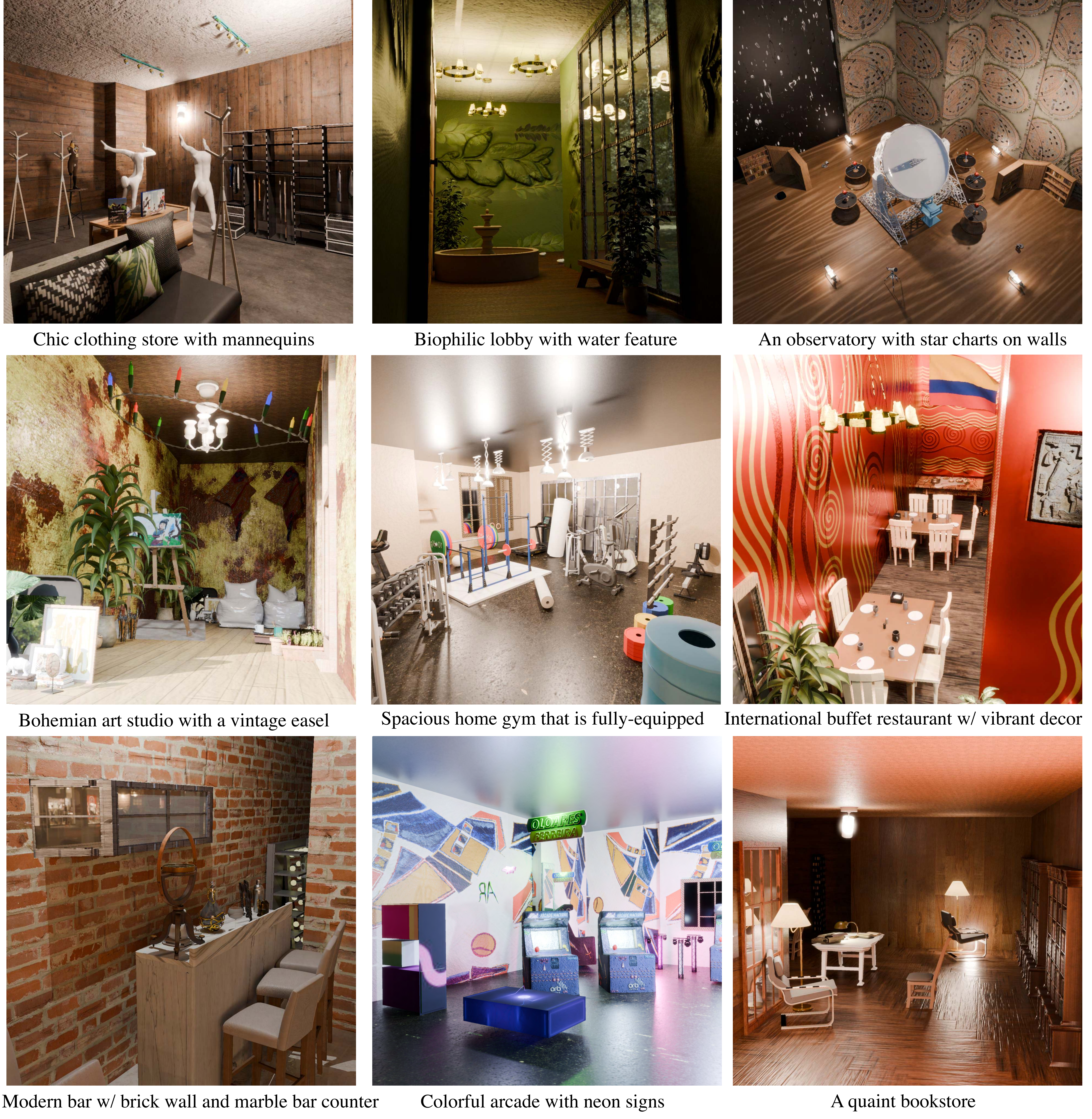}
    \caption{Sample 3D environments generated by \method, demonstrating control over assets, layout, material, and lighting.}
\end{figure}
}]

\input{sec/0_abstract} 

\input{sec/1_intro}

\begin{figure*}[!t]
  \centering
    \includegraphics[width=\linewidth]{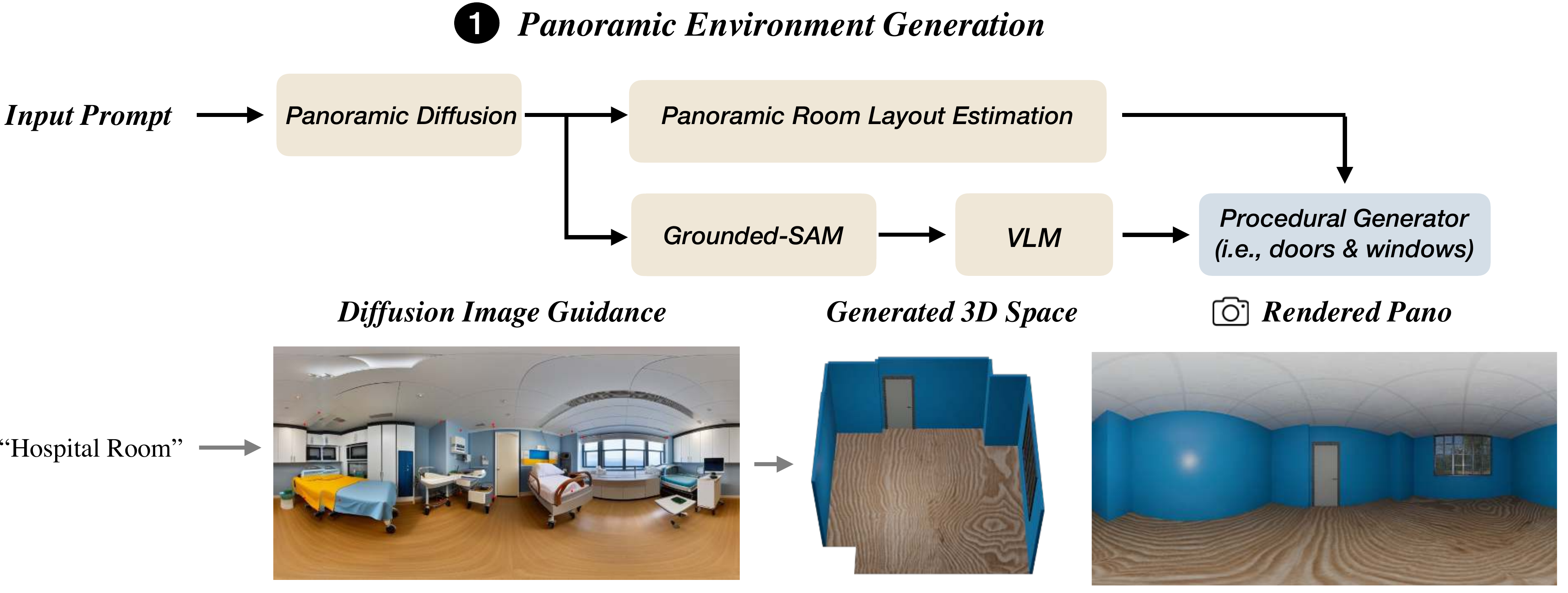} 
    \caption{\textbf{Overview of \method's \textit{\modulea}.} We use panoramic diffusion to generate a guiding 360° scene image, then extract the corners, windows, and doors information using a room layout estimation model, Grounded-SAM, and a VLM, respectively. These predictions are then used to construct the 3D room with fixtures procedurally.}
    \label{fig:modulea}
\end{figure*}

\input{sec/2_related}
\input{sec/3_method}

\input{sec/4_exp}
\section{Conclusion}
We introduce \textbf{3D-Generalist}, a framework that formulates text-driven 3D world creation as a sequential decision-making problem.
The framework can be integrated with image-to-3D reconstruction~\cite{sun2024partial,xu2024instantmesh}, image-to-PBR material generation~\cite{lopes2024material} or 3D asset generation models~\cite{nvidia2024edify3d}, enabling a fully generative graphics pipeline. Beyond content creation, the use of explicit 3D representations also facilitates generative simulation—enabling the automatic generation of environments for training RL policies~\cite{sun2024factorsim,wang2023robogen}.





{
    \small
    \bibliographystyle{ieeenat_fullname}
    \bibliography{main}
}
\input{sec/X_suppl}

\end{document}

%% file: preamble.tex
\usepackage[markup]{changes} 
\definechangesauthor[name=Yu-Hsin Chou, color=brown]{YHC}

\usepackage{xcolor} 

\usepackage{graphicx} 
\usepackage{float} 
\usepackage{array}    
\usepackage{multirow}

\usepackage{listings} 
\usepackage{mdframed} 

%% file: sec/0_abstract.tex
\begin{abstract}
Creating 3D graphics content for immersive and interactive worlds remains labor-intensive, limiting our ability to create large-scale synthetic data that can serve as training data to foundation models. Recent methods have been proposed to alleviate this, but they often focus on one particular aspect (e.g., layout) and fail to improve the quality of the generation through scaling computational resources. In this work, we recast 3D environment generation as a sequential decision-making problem, employing Vision-Language-Models (VLMs) as policies that output actions to jointly craft a 3D environment's layout, materials, lighting, and assets. Our proposed framework, \textbf{3D-Generalist}, trains VLMs to generate more prompt-aligned 3D environments via \textit{self-improvement} fine-tuning. We demonstrate the effectiveness of \textbf{3D-Generalist} and the proposed training strategy in generating simulation-ready 3D environments. Furthermore, we demonstrate its quality and scalability in synthetic data generation by pretraining a vision foundation model on the generated data. After fine-tuning the pre-trained model on downstream tasks, we show that it surpasses models pre-trained on meticulously human-crafted synthetic data and approaches results achieved with real data orders of magnitude larger.
\end{abstract}

%% file: sec/1_intro.tex
\vspace{-5mm}
\section{Introduction}
\label{sec:intro}


3D graphics design is vital in numerous fields, including gaming, augmented reality, virtual reality, and robotics simulation. However, creating graphics content such as immersive 3D worlds for entertainment or robotics simulation remains labor-intensive: artists and designers must select and create assets, apply materials, set up lights, and ultimately arrange all these aspects of an environment in conjunction. In this paper, we are interested in scaling up the generation of \textit{Simulation-Ready} 3D environments that can be readily used for downstream synthetic data applications or robotics task simulation, as opposed to 3D environments in representations such as NeRF or Gaussian Splats.

The advent of diffusion and large transformer models has introduced new ways to accelerate or automate specific components in a traditional graphics workflow. For example, Holodeck~\cite{yang2024holodeck} and RoboCasa~\cite{nasiriany2024robocasa} employ large language models (LLMs) to select, retrieve, and place 3D objects from asset libraries like Objaverse~\cite{deitke2023objaverse}, while URDFormer~\cite{chen2024urdformer} uses diffusion models to synthesize textures. However, these works are limited in their ability to scale with compute, leaving a gap between the generations and the quality demanded by downstream tasks. In this paper, we aim to create a system that can take open-ended text prompts as input and output 3D environments while leveraging more computational resources to enhance generation.\looseness=-1


To develop a method capable of scaling with computational resources, we posit that the key is the ability to self-correct and improve upon a model's own generation~\cite{kumar2024training,huang2023large}. This approach draws inspiration from the iterative refinement process of expert 3D artists, who progressively create detailed 3D worlds by rendering, inspecting, identifying mistakes, and repeatedly refining their designs with added details and corrections. Guided by this insight, we propose \method, a framework that unifies the generation of various aspects of a 3D environment --- materials, lighting, assets, and the layout --- by treating the problem as a sequential decision-making process. The idea is to employ large multi-modal models as \textit{action policies} guided by observations of the current 3D world.

\method comprises three modules: \textit{\modulea}, \textit{\moduleb}, and \textit{\modulec}. Key contributions include
\begin{itemize}[leftmargin=0em]
    \item We propose \textit{\modulea} to generate architectural layouts (i.e., floorplan, door and window placements) from text via a panoramic image-guided inverse graphics procedure.
    \item We propose \textit{\moduleb} which employs a VLM as a policy model that can jointly refine layout, materials, assets, and lighting of a 3D environment. We propose a self-improvement fine-tuning strategy that improves the \textit{\moduleb}'s ability to \textbf{self-correct} and craft more prompt-aligned environments iteratively, outperforming state-of-the-art baseline on generating simulation-ready 3D environments. We also show that the proposed self-improvement fine-tuning strategy further enhances the VLM's visual grounding abilities in general-domain benchmarks.
    \item We propose \textit{\modulec}, a second VLM policy capable of iteratively placing assets on top of other \textit{unlabeled} assets in a semantically aligned and physically plausible way.
    \item We demonstrate \method's ability to generate large-scale, high-quality synthetic data by training a vision transformer model that outperforms counterparts trained on carefully curated, manually crafted 3D environments, approaching results trained on a larger, real-world dataset.
\end{itemize}

%% file: sec/2_related.tex
\section{Related Work}
\subsection{Prompt-driven 3D Scene Generation}
Earlier forays into prompt-driven indoor scene synthesis rely on manual mapping between language and object placements such as Wordseye~\cite{coyne2001wordseye} and its follow-ups are symbolic rule-based and require significant human efforts to generalize to new domains and types of constraints~\cite{seversky2006real,chang2014learning,chang2017sceneseer,ma2018language,patil2024advances}.
Recent advances have explored two main directions. One line of work leverages diffusion prior using representations such as Neural Radiance Fields or Gaussian Splats~\cite{schult2024controlroom3d, zhou2024gala3d, po2024compositional, epstein2024disentangled, zhou2025dreamscene360}. These generated scenes lack separable, manipulable objects and surfaces, rendering them unsuitable for synthetic data applications where precise, instance-level annotations are required or robotics applications where robot-object interactions need to be simulatable.
Another line of research focuses on generating scenes that are \textit{simulation-ready}, often using intermediate representations (e.g., scene graphs or layouts) to arrange 3D assets from an asset repository \cite{rahamim2024lay, feng2024layoutgpt, yang2024holodeck, lin2024instructscene, hu2024scenecraft, yu2015clutterpalette, ccelen2024design, aguina2024open}.
Many of them focus on learning the distribution of assets and layout in scenes from a dataset~\cite{fisher2012example,paschalidou2021atiss,tang2024diffuscene}. Recently, the advent of Large Language Models (LLMs) and Vision-Language Models (VLMs) has enabled open-universe 3D scene synthesis, supporting the flexible generation of scenes without dependence on predefined labels or categories. For example, LayoutGPT~\cite{feng2024layoutgpt} prompts LLMs to directly generate 3D Layouts for indoor scenes. LayoutVLM~\cite{sun2024layoutvlm} uses VLMs to generate 3D layouts by generating objective functions that can be differentiable optimized. Unlike previous works that mostly focus on layout, \method jointly crafts layout for assets, materials, fixtures, and lighting.\looseness=-1







\subsection{Vision-Language Models for 3D Reasoning}
While (VLMs demonstrate promising general domain image understanding capabilities, they struggle to understand 3D or spatial relations~\cite{hong20233d, wang2024picture, chen2024spatialvlm}. Many works have sought to enhance VLMs' spatial understanding by incorporating techniques like visual markers. ``Visual Markers'' is a strategy to enhance VLMs’ ability to reason about spatial relationships by anchoring specific visual reference points in images~\cite{yang2023set}.
These markers help reduce reliance on textual cues alone, allowing VLMs to interpret spatial layouts and visual hierarchies within images directly. 
Others have curated and introduced new datasets to train the models on~\cite{goyal2020rel3d}.
For example, SpatialVLM \cite{chen2024spatialvlm} constructed a spatial reasoning dataset, enabling VLMs to perform tasks involving complex object placements and spatial configurations. SpatialRGPT~\cite{cheng2024spatialrgpt} further strengthens VLM spatial understanding by leveraging multi-view inputs to improve three-dimensional reasoning and object alignment across different perspectives. Similarly, \method also uses multi-view images with VLMs to perform spatial reasoning tasks and fine-tune the VLMs on task-specific data.\looseness=-1

%% file: sec/3_method.tex
\section{\method}
\method takes a text prompt as input and outputs a complete 3D room layout with materials, fixtures such as doors and windows, 3D assets, and their lighting configurations. Our idea is that creating a detailed and prompt-aligned 3D environment should be treated as a \emph{sequential decision-making} problem; thus, we iteratively refine scenes with our scene-level and asset-level policies. \method is a framework proposed around using Vision-Language Models (VLMs) to function as policies to optimize the 3D environments iteratively. Below, we outline this workflow and detail each of the three modules in our proposed framework.

\begin{figure*}[!t]
  \centering
   \includegraphics[width=\linewidth]{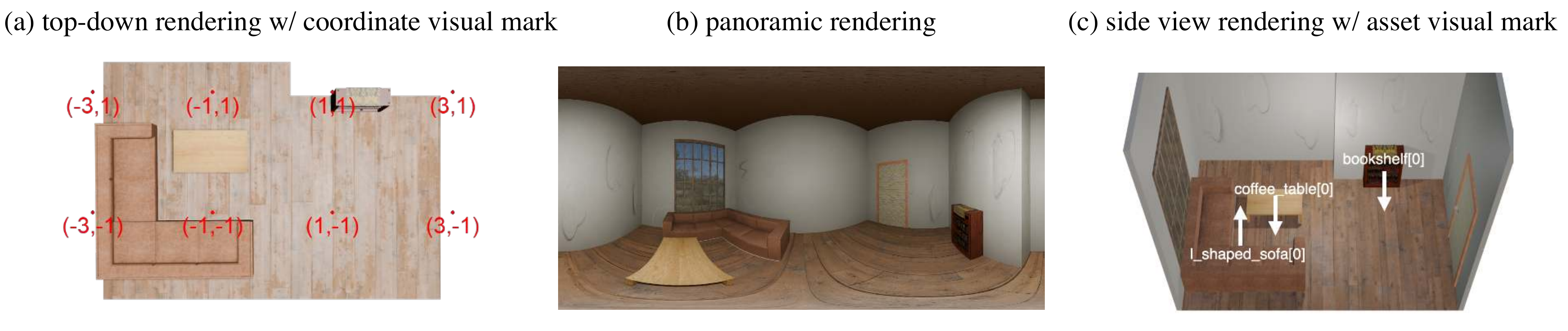} 
    \caption{\textbf{Overview of our Multi-View Representation \(\mathcal{I}_t\).} This illustrates how a 3D environment is rendered and fed as input to our VLM policy in \textit{\moduleb}. The first view is overlaid with visual marks of the x-y coordinate system of the current environment, the second view is a panoramic render of the room, and the third is overlaid with the variable names for the existing asset instances.}
    \label{fig:multi_view}
\end{figure*}

\subsection{\modulea} \label{sec:modulea}
This module, illustrated in Figure~\ref{fig:modulea}, initializes a base 3D room from the input text prompt, which includes the walls, floors, and fixtures including doors and windows. As opposed to LLMs predicting room coordinates, including door and window location, which often results in overly simplistic or unrealistic rooms, we propose to first use a panoramic diffusion model~\cite{feng2023diffusion360} to generate a 360$^\circ$ image as guidance; then, we use an inverse graphics procedure to construct the 3D environment detailed as follows:

\begin{enumerate}[leftmargin=1.6em] 
\item \textbf{Room Layout Estimation.} We take the panoramic image and use HorizonNet~\cite{sun2019horizonnet} model to derive the basic room structure (i.e., walls, floors, ceiling).
\item \textbf{Fixture Segmentation.} We apply Grounded SAM~\cite{ren2024grounded} to segment windows and doors. 
\item \textbf{VLM Annotation.} A Vision-Language Model (i.e., GPT-4o~\cite{achiam2023gpt}) inspects each segmented region to determine its type (e.g., \textit{single door}, \textit{double door}, \textit{sliding door}, or \textit{folding door}) and materials (i.e., \textit{door frame material}, \textit{door material}, and \textit{door knob material}). 
\item \textbf{Procedural Generation.} Subsequently, the rooms, doors, and windows are procedurally constructed in the corresponding 3D locations similar to \cite{deitke2022}. 
\end{enumerate}





\begin{figure*}[!t]
  \centering
   \includegraphics[width=\linewidth]{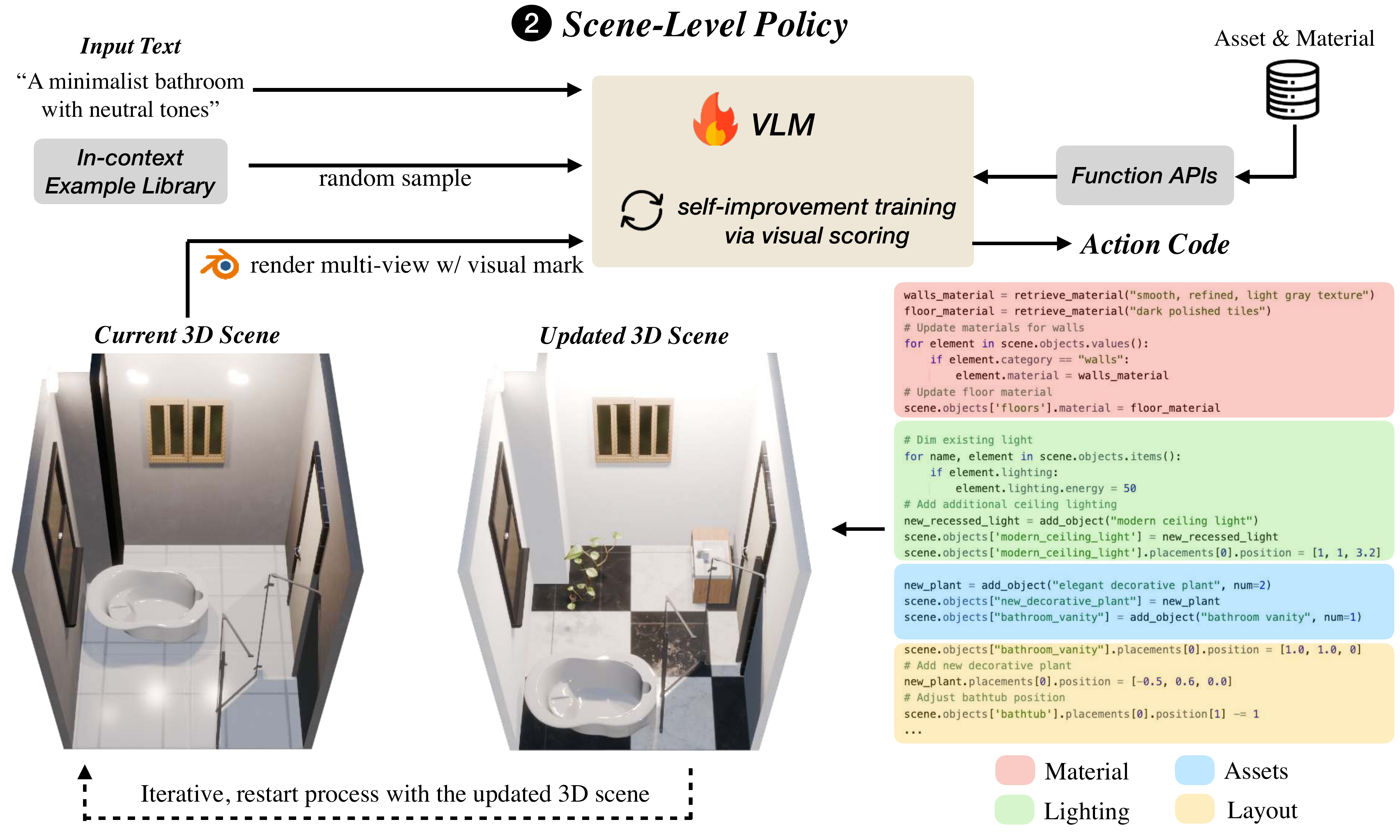}
    \caption{\textbf{Overview of \method's \textit{\moduleb}.} Starting with the current 3D scene, we render multiple views and combine the text prompt with an in-context example to guide a vision-language model (VLM) acting as an action policy. The VLM, during every round of the iterative process, generates a program that specifies updates in assets, materials, layout and lighting to the scene. We introduce a self-improvement training strategy to fine-tune our VLM to take actions that could lead to more prompt-aligned 3D scenes.}
    \label{fig:moduleb}
\end{figure*}

\subsection{\moduleb} \label{sec:moduleb}
This module adds assets to the initial 3D room generated by \textit{\modulea}. Models that can generate physically based rendering (PBR) materials or 3D assets from text or images produce outputs with significant artifacts. We opt to use an extensive repository of assets and materials instead of using diffusion models to generate images that inform the details of the 3D space \textit{\modulea}, as: (a) retrieving assets from a repository to match the objects in the image is empirically troublesome due to the lack of similar assets that can be reliably pose matched to the detected object in the image and (b) image-based detection or reconstruction methods are brittle under occlusions.
Our key insight is to build a model capable of \textbf{self-correction}. \textit{\moduleb} employs a VLM as the policy model, taking the current state of the 3D environment as input and outputting action in code to modify the 3D environment. Figure~\ref{fig:moduleb} illustrates this process. Below, we describe the key components.\looseness=-1

We use a domain-specific language (DSL) designed to represent 3D environments flexibly using a combination of code and natural language. Our Scene DSL defines key descriptors for scene element such as floors, walls, ceilings, objects, and lighting. The \textit{Category} descriptor specifies the type of element (\{\texttt{floors, walls, ceilings, objects}\}). The \textit{Placement} descriptor encodes spatial attributes, including position \( (x, y, z) \in \mathbb{R}^3 \), rotation \( \theta_z \in [0, 2\pi] \), and scale \( s \in \mathbb{R}^3 \), allowing multiple placements per element. The \textit{Material} descriptor provides a natural language description of surface properties. The \textit{Lighting} descriptor defines the type \( t \in \{\text{point}, \text{directional}, \text{area}\} \), intensity \( i \in \mathbb{R}^+ \), and color \( c = (r, g, b) \), where \( r, g, b \in [0,1] \) are normalized RGB values. More details can be found in the appendix.



\begin{figure*}[!t]
  \centering
   \includegraphics[width=\linewidth]{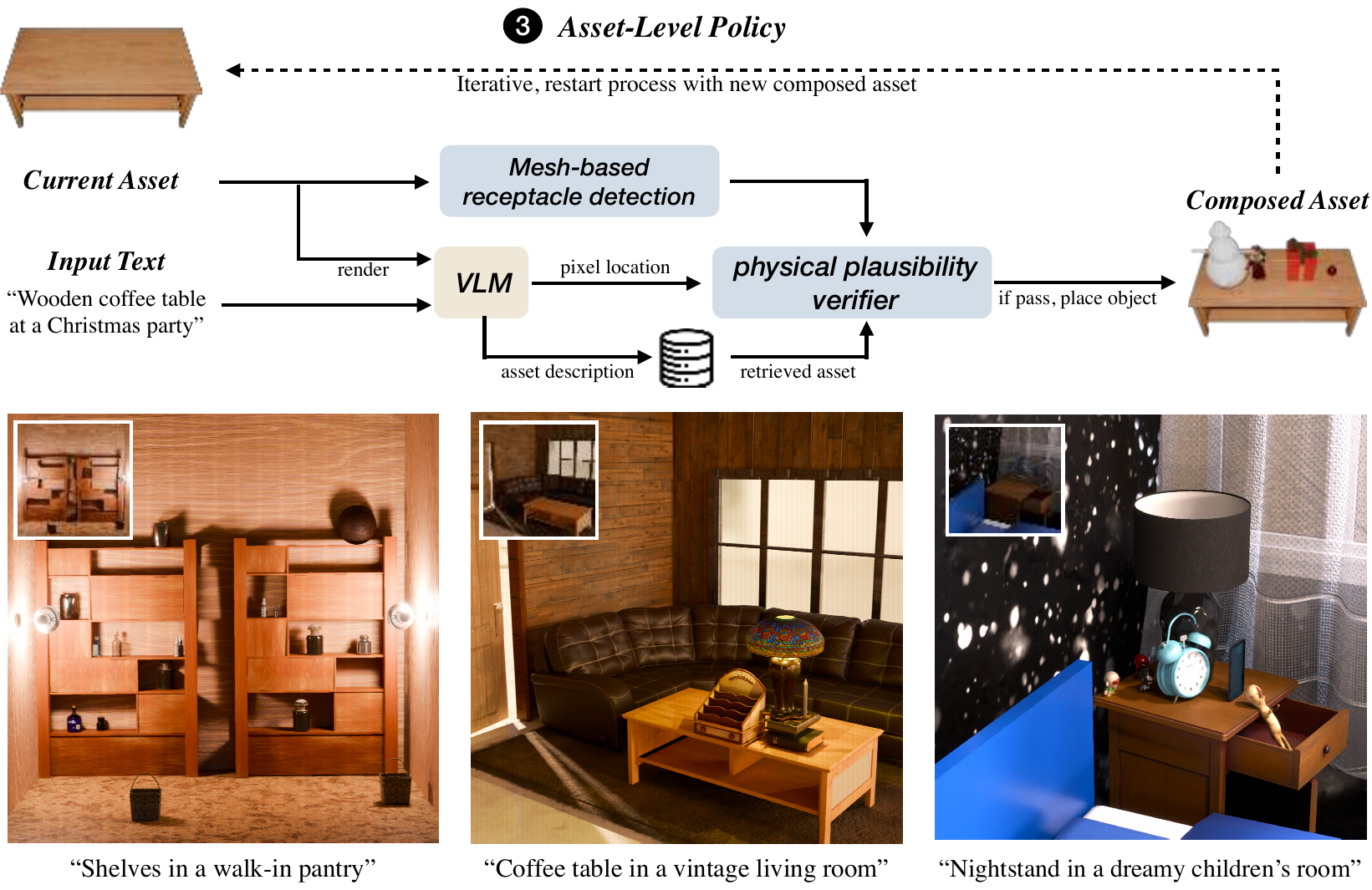} %
    \caption{\textbf{Overview of \method's \textit{\modulec}.} We qualitatively showcase its capability to handle diverse placement tasks, such as placing assets between shelves (i.e., as shown in the leftmost example), stacking assets on top of one another (i.e., the books and lamps as shown in the middle example), and placing assets in the open shelf of a nightstand (i.e., the figurine in the rightmost example).}
    \label{fig:modulec}
\end{figure*}
\paragraph{Vision-Language-Action Model}  
Let $S_0$ represent the 3D environment we obtained from \textit{\modulea}, \(S_t\) be the 3D environment at iteration \(t\), and let \(\mathcal{P}\) denote the input text prompt. Our VLM, \(\pi_\theta\), parameterized by \(\theta\), takes multi-view images \(\mathcal{I}_t\) of \(S_t\) (as shown in Figure~\ref{fig:multi_view}) along with \(\mathcal{P}\) as input and outputs an action code \(\displaystyle a_t \sim \pi_\theta\bigl(a \mid \mathcal{I}_t, \mathcal{P}\bigr)\). This action code is executed using exposed tools and function APIs—such as those for retrieving assets or materials based on natural language—which updates the 3D environment, expressed as \(\displaystyle S_{t+1} = f\bigl(S_t, a_t\bigr)\). Following the update, new images \(\mathcal{I}_{t+1}\) are rendered from \(S_{t+1}\) to serve as the next state representation. In Figure~\ref{fig:moduleb}'s example, the action code, upon execution, adjusts the position of the bathtub, adds a bathroom vanity and two plants, and increases the scene's lighting.\looseness=-1

\vspace{-2mm}
\paragraph{Self-Improvement Fine-tuning}  
Off-the-shelf VLMs often struggle to consistently generate action code that improves the resulting 3D environment. To address this, we propose a self-improvement fine-tuning strategy. Let \(\pi^{(i)}_{\theta}\) denote the VLM policy at the start of the \(i\)-th fine-tuning round. Each round begins with generating a set of tasks (i.e., text prompts). For each task, the 3D environment is initialized using \textit{\modulea}, and \(\pi^{(i)}_{\theta}\) iteratively updates the environment. At each state, multiple candidate actions are generated, and the top-scoring action sequences—those that yield the highest CLIP scores between the environment renderings and the input prompt—are retained. These top actions are then used to update the policy parameters via supervised fine-tuning, resulting in an improved policy \(\pi^{(i+1)}_{\theta}\). This process completes one self-improvement round, after which \(\pi^{(i+1)}_{\theta}\) can generate new tasks and environments for further refinement in subsequent rounds. Additionally, we introduce an \emph{in-context library}, a collection of action codes that significantly enhances CLIP alignment scores. Empirically, we find that leveraging the \emph{in-context library} during data generation substantially improves performance by promoting diversity and increasing the likelihood of discovering effective action sequences. We use GPT-4o~\cite{achiam2023gpt} as the base model and leveraged its publicly available fine-tuning API in our experiments. More details can be found in the supplementary material.

\subsection{\modulec} 
Our scene-level VLM policy, used in \textit{\moduleb}, often omits smaller objects (e.g., books, plates, utensils) and focuses on larger and more defining environmental assets. In contrast, placing small assets requires a different design approach to ensure strong physical plausibility.
To address this, we introduce \textit{\modulec}, which refines the environment by composing smaller assets with “receptacle objects” (e.g., shelves, tables, counters), closely mimicking the construction process of physically grounded environments. \textit{\modulec} is equipped with physical plausibility verifiers applied prior to placement. The system is inherently extensible, supporting custom plug-in verifiers such as physics engines or logic-based constraints. The process begins with GPT-4o determining whether a 3D asset qualifies as a ``receptacle object'' that can host smaller items. Once identified as a receptacle, we initiate an iterative process to place these smaller assets. Before the first round, we have a base object \(\mathcal{O}\) and a text prompt \(\mathcal{P}\) (``wooden office desk in a writer's home''). Below, we describe the process at the $k$-th round: 

\paragraph{Mesh-based Placeable Surface Detection.} We use a mesh-based surface detection approach to find valid surfaces on the receptacle object (or on previously placed items, allowing for stacking). See details in supplementary. We then render the receptacle object from a randomly sampled angle on a hemisphere, generating an image rendering. The camera angle is sampled at each round as that exposes different placeable surfaces across iterations.

\paragraph{VLM as a Placement Policy.} This rendered image \(\mathcal{I}'_k\) (distinct from the multi-view renderings \(\mathcal{I}_k\) in \textit{\moduleb}) and \(\mathcal{P}\) are fed into a separate policy model \(\pi'_\phi\), a pre-trained VLM from \cite{deitke2024molmo}. Here, we chose \cite{deitke2024molmo} because it excels in pixel-level precision required in asset placement, distinct from the choice of GPT-4o in \textit{\moduleb} policy that requires strong semantic reasoning.
Formally, the action chosen in round $k$ can be formally denoted as $a_k' \sim \pi'_\phi\bigl(a \,\mid\, \mathcal{I}'_k,\;\mathcal{P}\bigr)$. Here, an action is represented as \(a'_k = (o'_k, p'_k)\), where \(o'_k\) denotes the pixel location in the rendered image indicating the placement position, and \(p'_k\) represents the text description used to retrieve 3D assets from the repository.

\paragraph{3D Placement and Environment Update.} Using the pixel location \(o'_k\) generated by \cite{deitke2024molmo} and associated camera parameters, we generate a 3D ray to identify a precise placement point on the mesh. If the location intersects one of the previously identified valid surfaces, we retrieve the asset specified by \(p'_k\). After verifying that this asset can be realistically placed (via mesh collision checks), we place then repeat this process with the newly composed asset.  By iterating over these steps, \textit{\modulec} which recursively introduces the finer details—often overlooked by scene-level policies—into the 3D environment e.g., a book can be placed on the table, then a pen can be placed on the book.\looseness=-1

%% file: sec/4_exp.tex
\section{Experiments}
We evaluate \method through experiments aimed at answering the following questions: 
\begin{description}
    \setlength{\itemsep}{0pt} 
    \setlength{\parsep}{0pt} 
    \setlength{\partopsep}{0pt} 
    \item[(a)] How does \method compare to existing methods on generating simulation-ready 3D environments? 
    \item[(b)] Does our proposed self-improvement fine-tuning strategy enable \textit{\moduleb} iteratively refine the 3D environments to be more prompt-aligned?
    \item[(c)] Can our proposed \textit{\modulec} effectively place small objects in a semantically coherent manner?
    \item[(d)]  Does our method enable effective scaling of 3D data generation for training robust visual feature extractors?
\end{description}

\begin{figure*}[!th]
  \centering
   \includegraphics[width=\linewidth]{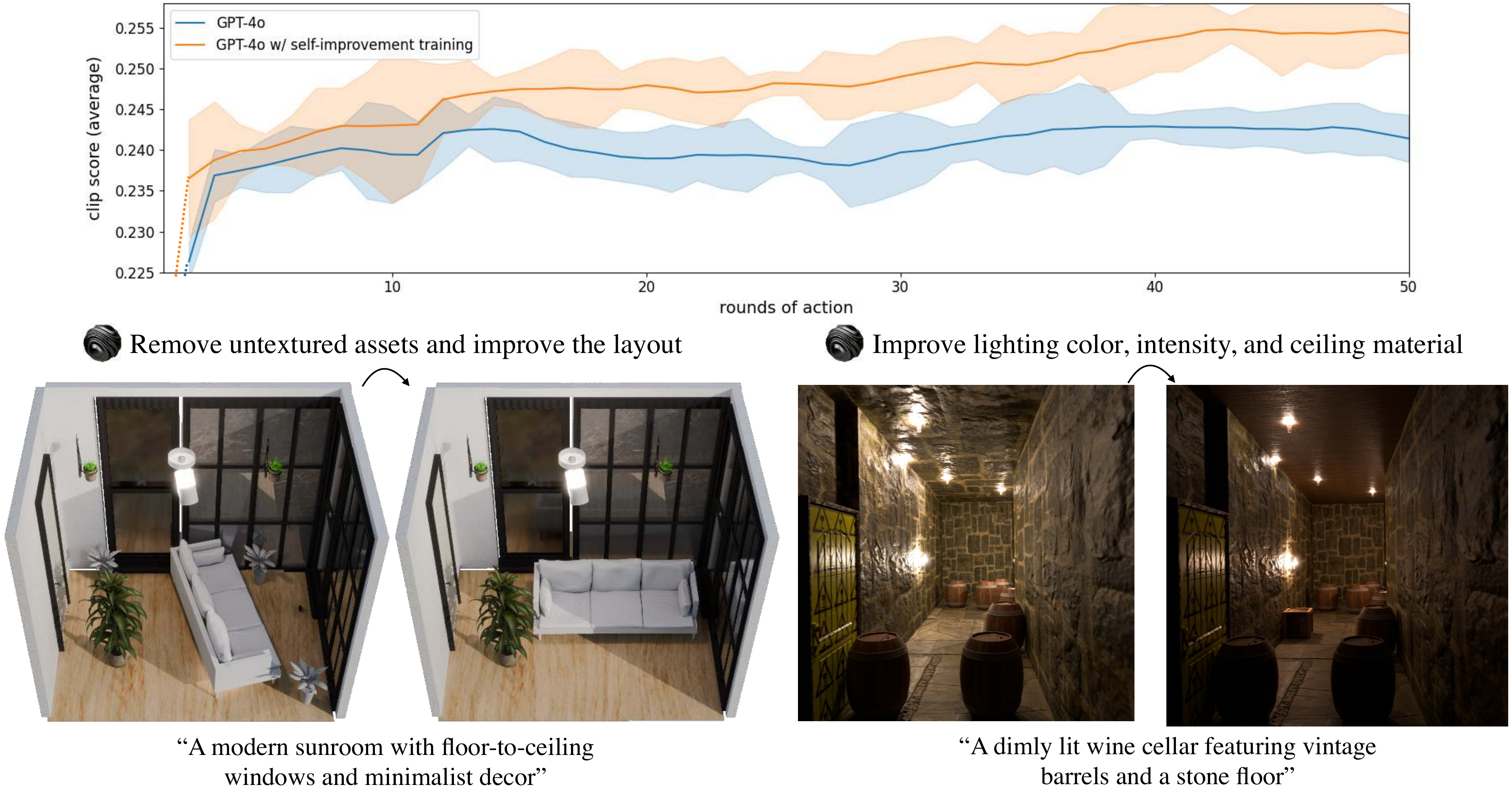}
    \caption{\textbf{Evaluation of the self-improving fine-tuning strategy.} The top graph shows averaged results, indicating that GPT-4o improves 3D prompt alignment iteratively only after self-improvement fine-tuning. The bottom examples highlight \method's self-correction: refining untextured assets and layout in one case, and adjusting lighting and ceiling material in another.}
    \label{fig:self_correct}
\end{figure*}

\subsection{Simulation-Ready 3D Environment Generation}
Following existing works~\cite{radford2021learning,hollein2023text2room,zhou2024gala3d,zhang2023scenewiz3d,sun2024layoutvlm}, we evaluate simulation-ready 3D environments by measuring physical plausibility and semantic coherence. Physical plausibility is measured using the \textit{Collision-Free Score (CF)} and \textit{In-Boundary Score (IB)}. All assets are enforced to be placed, with remaining assets randomly placed if a method fails. Semantic coherence is assessed using \textit{Positional Coherency (Pos.)} and \textit{Rotational Coherency (Rot.)}, measuring alignment with the input prompt. To evaluate semantic coherence across layouts without groundtruth, we use \textit{GPT-4o} to score layouts based on top-down and side-view renderings and the language instructions.\textit{Physically-Grounded Semantic Alignment Score (PSA)} is calculated simply the GPT-4o rating weighted by physical plausibility. Scores range from 0 to 100, with higher scores indicating better performance. Note that we don't enable gravity in all experiments, and that a collision is considered to occur when the intersecting area between two meshes exceeds 0.1 m² in Blender.

\input{tables/layout}
\input{tables/self_improvement}

\paragraph{Comparative Study}
To answer (a), we choose LayoutGPT~\cite{feng2024layoutgpt}, Holodeck~\cite{yang2024holodeck}, and LayoutVLM~\cite{sun2024layoutvlm}, the state-of-the-art in \textit{sim-ready} 3D environment generation methods as baselines. While there are many other works on 3D scene generation like Physcene~\cite{yang2024physcene} and DiffuScene~\cite{tang2024diffuscene}, they are either limited to a fixed set of object categories (i.e., not \textit{open-vocabulary} methods) or rely on representations such as Gaussian or NeRF as opposed to resorting to reasoning models (i.e., LLMs, VLMs) with object-centric representation.
\method outperforms all baselines using the same set of prompts used in Holodeck~\cite{yang2024holodeck} (over 250 prompts across 50 room types). 



\paragraph{\moduleb Evaluation}
To answer (b), we assess our self-improving fine-tuning strategy using quantitative metrics and qualitative analysis (CLIP scores in Table~\ref{tab:self_improvement_ablation} and Figure~\ref{fig:self_correct}). 
The x-axis in Figure~\ref{fig:self_correct} represents "rounds of action" within a generation sequence. While GPT-4o (baseline) shows some early refinement, our fine-tuned model demonstrates significantly stronger iterative improvement. Qualitative examples highlight \textit{\moduleb}’s ability to self-correct.

\paragraph{Additional Ablation on Fine-tuning}
We explore how self-improvement fine-tuning impacts VLM performance in general domains. Since our approach trains the VLM to generate better action code from image observations, we hypothesize it enhances visually grounded understanding, transferable to general-domain imagery. To verify this, we evaluate \method's \textit{\moduleb} on two commonly used visual hallucination benchmarks (i.e., \textbf{Object HalBench}~\cite{rohrbach2018object} and \textbf{AMBER}~\cite{wang2023llm}) that directly assess the accuracy of the VLM's visual grounding capabilities (details in supplementary). 

Results in Table~\ref{tab:visual_hal} show that the base VLM GPT-4o benefits significantly from our self-improvement fine-tuning, with the fine-tuned \method exhibiting lower hallucination rates across both benchmarks and multiple metric levels, except for a slight decrease in object coverage (COVER) on AMBER. These results demonstrate that, in addition to generating 3D scenes more aligned with the prompt, our self-improvement fine-tuning also benefits VLMs' general-domain visual understanding, effectively reducing their visual hallucinations.

\begin{figure*}[!t]
    \centering
    \includegraphics[width=\linewidth]{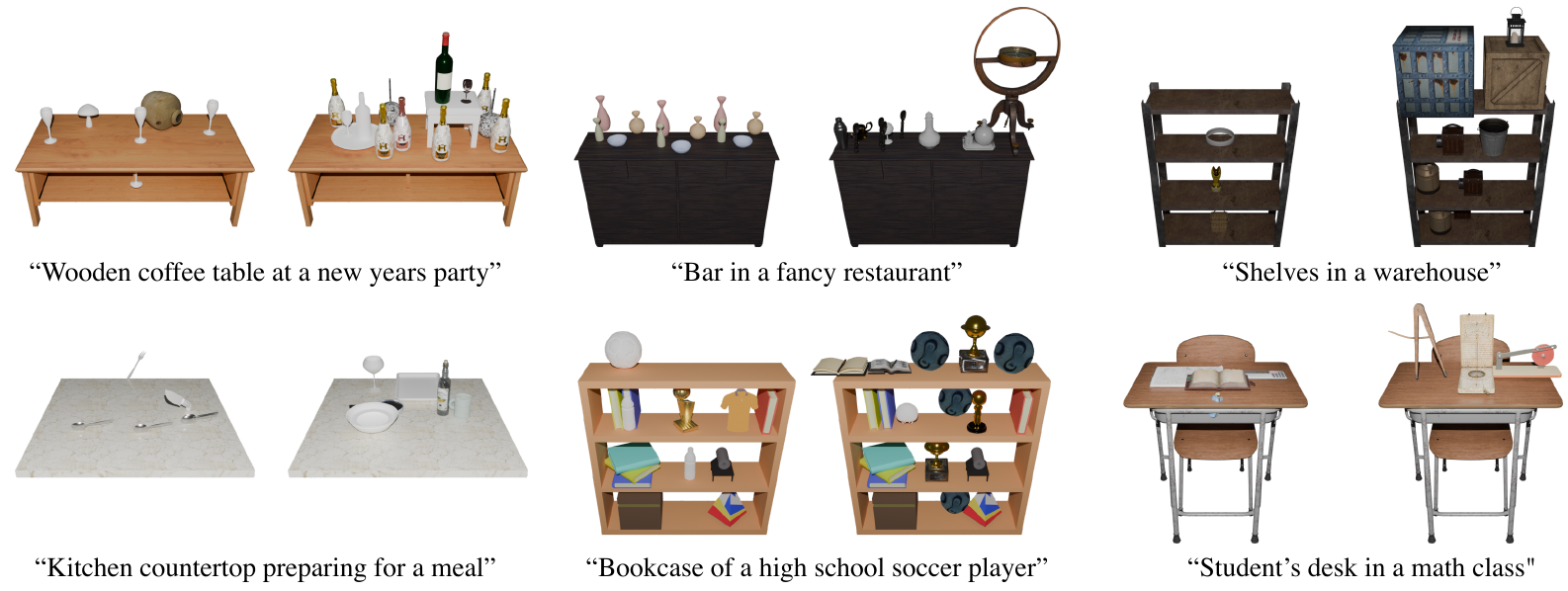} 
    \caption{Comparison between the baseline inpainting method (left) and \method's \textit{\modulec} (right).}
    \label{fig:inpaint}
\end{figure*}

\paragraph{\modulec Evaluation}
In our experiments, GPT-4o is responsible for determining whether a 3D asset qualifies as a receptacle object, 
while another VLM, Molmo-7B~\cite{deitke2024molmo}, provides the visual cues for pixel-level placement. 
As shown in Figure~\ref{fig:modulec}'s qualitative examples, \textit{\modulec} can place books on shelves, stack plates on tables, or fill open shelves with suitable items.

To answer \textbf{(c)} quantitatively, we ran the experiment on 15 different ``receptacle assets'' split equally into three categories: small tables (nightstand, student desk, bar, side table, foyer table), large tables (coffee table, glass coffee table, kitchen countertop, metal table, office desk), and multi-level assets (bookshelf - open back, bookshelf - closed back, pantry shelf, work shelf, warehouse shelf). After finishing placement, we render from four equally spaced camera positions around the object and report the maximum CLIP similarity score between the rendered image and the input text prompt. For \textit{\modulec}, we set the maximum number of successful asset placements to 10.

\input{tables/visual_hal}

Across the 15 test cases, the base receptacle objects (without any other objects) have an average CLIP score of 0.264. The baseline method produces composed assets with an average CLIP score of $0.269 \pm 0.001$, and our \textit{\modulec} produces an average score of \boldmath{\textbf{$0.282 \pm 0.001$}}
, demonstrating the ability to place assets on ``receptacle assets'' in a prompt-aligned and physically plausible way~\footnote{During our evaluation process, an asset can only be placed if the resulting scene is physically plausible (i.e., no mesh collision).} Qualitatively, we observe that the baseline inpainting-based method regularly generates many overlapping objects and cannot place an item behind another. Our method, however, has access to multiple views during the iterative process and can thus spread out the object placement more naturally. Refer to Figure~\ref{fig:inpaint} and the supplementary material for qualitative comparisons that showcase of the importance of leveraging the semantic knowledge in the VLM (i.e., Molmo) to predict asset placement.



\input{tables/florence}
\footnotetext{The training GPU hours for the original Florence-v2 is estimated.}
\subsection{Downstream Applications}
To demonstrate \method's capability in scaling up 3D data, we conduct large-scale pretraining of vision encoders on the generated synthetic data. We render millions of images from generated scenes using Omniverse to create a large-scale dataset and compare feature extractors trained on our synthetic data with those trained on HyperSim~\cite{hypersim}, an artist-crafted dataset, along with a real-world dataset.

In our experiments, we first train a vision encoder using the Florence 2~\cite{florence2} framework with a DaViT-B~\cite{florence2} backbone. Next, we initialize an image classification model from the pretrained checkpoint and finetune the model on ImageNet1K~\cite{deng2009imagenet}. The only variable in our experiments is the selection of the pretraining dataset. We convert Hypersim into Florence 2 format with three tasks: 2D object detection, semantic segmentation, and bounding box to segmentation. Doing so results in just over 860K labels, and we also generate a dataset consisting of the same number of labels using \method. Finally, as \method has the advantage of scaling datasets with ease, we generate an additional 11.5M labels to perform pretraining with. 

All results are provided in Table~\ref{tab:downstream_task_results} and compared to finetuning on the publicly available DaViT-B weights from Florence 2, which is trained on 5B real-world labels. All experiments are conducted on a 96×H100s cluster, with the runs below 1 million labels taking roughly 96 hours and the larger run of 12 million labels taking 200 hours. With commonsense and visual spatial reasoning of VLMs, \method produces high-fidelity scenes that better reflect real-world object distribution and layouts. This leads to more effective transfer on real-world downstream tasks. Furthermore, our synthetically generated environments can effectively scale up training data for vision models, achieving comparable or superior performance to manually curated datasets such as Hypersim while significantly reducing data collection effort.

%% file: tables/layout.tex
\begin{table}[t]
\label{tab:layout}
    \caption{Physical plausibility and semantic alignment, using the same metrics and setting as Tab. 2 of LayoutVLM \cite{sun2024layoutvlm}.  All scores range from 0 to 100 ($\uparrow$ is better). Due to the iterative nature, perfect physical plausibility can be achieved as the model can self-correct asset placement over rounds of iteration.}
    \centering
    \resizebox{.8\columnwidth}{!}{%
    \begin{tabular}{lccccc}
    \toprule
     & \multicolumn{2}{c}{Physics} & \multicolumn{2}{c}{Semantics} & \multicolumn{1}{c}{Overall Score} \\
    \midrule
    & CF & IB & Pos. & Rot. & PSA\\
        \cmidrule(lr){2-3}
    \cmidrule(lr){4-5}
    \cmidrule(lr){6-6}
   LayoutGPT & 83.8 & 24.2 & \textbf{80.8} & 78.0 & 16.6 \\
   Holodeck  & 77.8 & 8.1 & 62.8 & 55.6 & 5.6 \\
   I-Design & 76.8 & 34.3 & 68.3 & 62.8 & 18.0 \\
   LayoutVLM & 81.8 & 94.9 & 77.5 & 73.2 & 58.8 \\
   \midrule
    \method & \textbf{99.0} & \textbf{98.0} & 78.2 & \textbf{79.1} & \textbf{67.9} \\
    \bottomrule
    \end{tabular}
    \label{tab:physplaus}
    }
\end{table}

%% file: tables/self_improvement.tex
\begin{table}[t]
    \centering
    
    \caption{Comparative Analysis and Ablation Study on \method's \textit{\moduleb}. ``\# Training Runs'' refer to the total number of times we perform (self-improvement) fine-tuning on the VLM, and ``\# Actions'' refer to the number of actions taken by the VLM in \textit{\moduleb} per scene generation. We report the CLIP score of the final generated scene.}
    \resizebox{\columnwidth}{!}{%
        \begin{tabular}{lccc}
        \toprule
        \textbf{Method} & \textbf{\# Training Runs} & \textbf{\# Actions} & \textbf{CLIP Score} \\
        \midrule
        Random noise (lower bound) & \large{-} & \large{-} & \large{0.026} \\
        LayoutGPT & \large{-} & \large{-} & \large{0.228} \\
        Holodeck & \large{-} & \large{-} & \large{0.231} \\
        LayoutVLM &  \large{-} & \large{-} & \large{0.239} \\
         \method  & \large{3} & \large{10} & \textbf{\large{0.275}} \\\midrule
        \method w/o fine-tuning & \large{0} & \large{10} & \large{0.252} \\
        \method w/o in-context library &  \large{0} & \large{10} & \large{0.237}  \\ \midrule 
        \method ablations  & \large{3} & \large{3} & \large{0.254} \\
                                  & \large{2} & \large{3} &  \large{0.251} \\
                                 & \large{1} & \large{3} &  \large{0.248} \\
                                  & \large{0} & \large{3} &  \large{0.242} \\
                                  & \large{0} & \large{0} & \large{0.159} \\
        \bottomrule
        \end{tabular}
    }
    \label{tab:self_improvement_ablation}
    \vspace{-2mm}
\end{table}

%% file: tables/visual_hal.tex
\begin{table}[t]
    \centering
    \caption{
    Our self-improvement fine-tuning strategy demonstrates positive transfer that effectively reduces VLM's visual hallucinations in general-domain images.}
    \resizebox{\columnwidth}{!}{%
    \begin{tabular}{ccccccc}
        \toprule
         & \multicolumn{2}{c}{\textbf{Object HalBench}} & \multicolumn{4}{c}{\textbf{AMBER-Generative}} \\
         \cmidrule(lr){2-3} \cmidrule(lr){4-7}
         \textbf{Method}
         & CHAIRs$\downarrow$
         & CHAIRi$\downarrow$
         & CHAIR$\downarrow$
         & COVER$\uparrow$
         & HAL$\downarrow$
         & COG$\downarrow$
         \\
        \midrule
        \multirow{2}{*}{\shortstack{\method \\(GPT-4o)}} 
            & \multirow{2}{*}{\large{10.3}}
            & \multirow{2}{*}{\large{5.4}}
            & \multirow{2}{*}{\large{3.3}}
            & \multirow{2}{*}{\textbf{\large{61.8}}}
            & \multirow{2}{*}{\large{16.5}}
            & \multirow{2}{*}{\large{0.8}} \\
            & & & & & &  \\
        \hline
        \multirow{2}{*}{\shortstack{\method \\(GPT-4o \textbf{finetuned})}} 
            & \multirow{2}{*}{\textbf{\large{7.7}}}
            & \multirow{2}{*}{\textbf{\large{4.6}}}
            & \multirow{2}{*}{\textbf{\large{3.2}}}
            & \multirow{2}{*}{\large{60.8}}
            & \multirow{2}{*}{\textbf{\large{15.7}}}
            & \multirow{2}{*}{\textbf{\large{0.7}}} \\
            & & & & & &  \\
        \bottomrule
    \end{tabular}
    }
    \vspace{-1mm}
    \label{tab:visual_hal}
\end{table}

%% file: tables/florence.tex
\begin{table}[t]
    \centering
    \caption{Downstream application: Pretraining Florence-v2 on generated datasets and reporting fine-tuned ImageNet performance.}
    \resizebox{0.9\columnwidth}{!}{%
        \begin{tabular}{ccc}
            \toprule
            \textbf{Pretraining Dataset} & \textbf{\# Labels} & \textbf{ImageNet-1K Top 1 \(\uparrow\)} \\ 
            \midrule
            Hypersim & 861,080 &  0.727 \\
            \method & 861,080 & 0.731 \\
            \method & 12,175,588 & 0.776 \\
            Florence 2 (real) & 5,000,000,000 & 0.786 \\
            \bottomrule
        \end{tabular}
    }   
    \label{tab:downstream_task_results}
\end{table}

%% file: sec/X_suppl.tex
\clearpage
\setcounter{page}{1}
\maketitlesupplementary

\section{Additional Details of \method}
Here, we provide more detail for the implementation of various components in \method. 

\subsection{\moduleb}
The base prompt we use as input to our VLM is:
\begin{lstlisting}[]
**Task**: As a programmer, you are required to complete an implementation. Use a Chain-of-Thought approach to break down the problem, create pseudocode, and then write the code in Python language. Ensure that your code is efficient, readable, and well-commented.
Return the requested information from the function you create. Remember to call the function your create towards the end.

**Dos and Don'ts**:
* Only use the API functions provided. Please do not import any of them. You can use the functions directly, assuming they are already imported.
* Give scene elements unique and descriptive names. For example, brown_leather_sofa, ceiling_light, tall_floor_lamp.
* You can duplicate or remove assets and lights by manipulating the attribute "placements" of the asset instance or just delete the entry from the scene.objects dictinoary.
* Make sure to provide detailed but key-word focused descriptions of any new objects or materials you aim to retrieve. For example, "L-shaped, brown leather sofa with plush seat and colorful pillow" instead of "a sofa".
* Positions are bounding box center in meters, given as [x, y, z]. Rotations are in degrees. The assets are upright, so you only have to change and specify the one number: z-axis rotation.
* Make all modifications to the SceneDef instance named "scene" as we will read out the results from the variable named `scene`.
* Do not re-initialize the scene by calling SceneDef() again. The scene is already initialized and you can access the current scene using the variable "scene".
* Note that the lighting and material attributes are by default None. You need to set them by instantiating the class Lighting or Material if you want to change the lighting or material of the scene element.
* Make sure to always specify the position and rotation of newly added objects, otherwise the objects will be added at the origin.

**3D Convention:**
- We use a right-handed coordinate system.
- For the scene, the X-Y plane is the floor, the Z axis points up to the ceiling.
- For each asset, the front face points in the positive X axis. The Z-axis points up. For example, a z-axis rotation of 90 means that the object will be rotated to face the positive Y axis.
- The bounding box of the asset is aligned with the asset's local frame. The local origin is at the center in the X-Y plane and the bottom of the asset in the Z axis. An object having a z-axis position of 0 means that object is on the floor.
\end{lstlisting}

\paragraph{Scene Domain Specific Language}
Here, we detail the types of descriptors defined in our Scene DSL.
The \textit{Category} descriptor, \( C \), specifies the type of the scene element, where \( C \in \{\text{floors}, \text{walls}, \text{ceilings}, \text{objects}\} \). The \textit{Placement} descriptor, \( P = [(x, y, z), \theta_z, s] \), defines the spatial attributes of the element, where \( (x, y, z) \in \mathbb{R}^3 \) specifies the position in 3D space, $\theta_z \in [0, 2\pi]$ represents the rotation around the z-axis, and \( s \in \mathbb{R}^3 \) is the scale of the element. Placement can include a list of such tuples to allow for multiple placements of a particular scene element. The \textit{Material} descriptor, \( M = \text{desc} \), provides a natural language description (\( \text{desc} \in \text{NL} \)) of the element’s surface properties. The \textit{Lighting} descriptor, \( L = (t, i, c) \), specifies the type of the light, \( t \in \{\text{point}, \text{directional}, \text{area}\} \), along with its intensity \( i \in \mathbb{R}^+ \) and color \( c \in \mathbb{R}^3 \), where \( c = (r, g, b) \) and \( r, g, b \in [0, 1] \) represent the normalized RGB color values. Finally, the \textit{Metadata} descriptor, \( D \), captures additional category-specific attributes, details of which are provided in the supplementary materials. Below, we provide the implementation in Python:
\begin{lstlisting}[]
from pydantic import BaseModel, Field, conint
from typing import List, Optional, Dict, Literal, Any
import math
import numpy as np

class Placement(BaseModel):
    position: List[float] = Field(description="(x, y, z) position of the asset. z=0 means the asset is on the ground.", default=[0, 0, 0])
    rotation: List[float] = Field(description="(x, y, z-axis) Rotation of the asset in degrees. Most assets are upright in the source data, so rotation is mostly around the z-axis.", default=[0, 0, 0])
    scale: float = Field(description="Axis-aligned size of the bounding box before the rotation is applied", default=1.)

class Material(BaseModel):
    id: str = Field(description="ID of the material")
    description: str = Field(description="Description of the material")

class Lighting(BaseModel):
    is_ceiling_light: bool = Field(description="Whether the lighting is a ceiling light", default=False)
    light_type: Literal["spotlight", "directional", "point"] = Field(description="Type of the light", default="point")
    energy: float = Field(description="Energy / intensity of the light in lux (indoor lights typically range from 100 to 1000)", default=100)
    color: List[float] = Field(description="Color of the light in RGB format (0-1, not 0-255)", default=[1, 0.75, 0.6])

class SceneElement(BaseModel):
    description: str = Field(description="Detailed natural language description of the scene element", default="")
    category: Literal["floors", "walls", "ceilings", "doors", "windows", "objects"] = Field(description="Category of the scene element")
    placements: List[Placement] = Field(description="Placement information for the scene element")
    bbox_size: Optional[List[float]] = Field(description="Axis-aligned size of the bounding box before the rotation is applied", default=None)
    material: Optional[Material] = Field(description="Default material applied to the scene element (applicable for floors, walls, ceilings, doors, windows)", default=None)
    lighting: Optional[Lighting] = Field(description="Lighting attached to the scene element that emits light (only applicable for 'objects' category)", default=None)
    metadata: Optional[Dict[str, Any]] = Field(description="Additional metadata for the scene element", default=None)

class SceneDef(BaseModel):
    objects: Dict[str, SceneElement] = Field(description="Dictionary of assets in the scene with asset variable name as keys. The asset variable name should be a valid Python variable name.", default={})
    room_type: Optional[str] = Field(description="room type", default="an interior scene")

scene = SceneDef()
\end{lstlisting}

\paragraph{Asset-level policy details}
After mesh-based surface detection, we find valid surfaces on the receptacle object by identifying all planes with normals deviating from [0, 0, 1] by less than 10 degrees then cluster them together and keep the clusters above a threshold area. 

\paragraph{Functional APIs}
We provide the set of functions that we provide to our VLM in \textit{\moduleb}, mainly \textbf{retrieve\_material} to retrieve materials and \textbf{add\_object} to retrieve 3D assets from language.

\begin{lstlisting}[]
This is the documentation for the functions you have access to. You may call any of these functions to help you complete the task.

retrieve_material(description: str) -> scene_definition.Material:
'retrieve_material' is a tool that can retrieve a PBR material from language description.
    It returns a dictionary containing the id and description of the retrieved material.

    Parameters:
        description (str): The description of the material to change the wall to

    Returns:
        Material: A dictionary containing the id and description of the retrieved material.

add_object(description: str, bbox_size: list = None, num: int = 1) -> scene_definition.SceneElement:

    'add_object' is a tool that can retrieve an asset from the database from language description.
    It is necessary to call this function to add any new object to the scene, including lighting related objects.
    It returns a dictionary containing the id, description, size, and other optional metadata of the retrieved asset.
    Note that the asset is added to the origin (center of the scene) with no rotation so you might need to adjust the position and rotation of the retrieved asset.

    Parameters:
        description (str): The description of the asset to retrieve
        bbox_size (list, optional): A list of 3 floats [x, y, z] specifying the desired dimensions of the object in meters.
                                   This controls the size of the generated object. The z-axis value denotes the height of the object. Defaults to [1.0, 1.0, 1.0].
        num (int, optional): Number of instances of this object to create. Defaults to 1.

    Returns:
        SceneElement: a scene element containing the id, description, size, and other optional metadata of the retrieved asset.

load_image(image_path: str) -> numpy.ndarray:
'load_image' is a utility function that loads an image from the given file path string.

    Parameters:
        image_path (str): The path to the image.

    Returns:
        np.ndarray: The image as a NumPy array.
\end{lstlisting}

\paragraph{In-Context Library}
\noindent
Our in-context library is a curated collection of code snippets that have demonstrated at least a 10\% improvement in CLIP-alignment metrics when modifying a 3D environment. These snippets serve as in-context examples during generation. By randomly sampling from this library when producing candidate actions, the policy is nudged toward more diverse and effective action codes. After each self-improvement training round, newly discovered high-quality prompt and action code pairs are appended to this library. Below, we showcase an example code snippet from our in-context library for the prompt: \textit{a chic hair salon with round mirrors, pink chairs, and pot plants}.
\begin{lstlisting}[]
# Add pink salon chairs
pink_salon_chairs = add_object("plush pink salon chair", num=4)
scene.objects['pink_salon_chairs'] = pink_salon_chairs

# Position the chairs along a wall
scene.objects['pink_salon_chairs'].placements[0].position = [-1.5, 1.5, 0]
scene.objects['pink_salon_chairs'].placements[0].rotation = [0, 0, 0]
scene.objects['pink_salon_chairs'].placements[1].position = [-0.5, 1.5, 0]
scene.objects['pink_salon_chairs'].placements[1].rotation = [0, 0, 0]
scene.objects['pink_salon_chairs'].placements[2].position = [0.5, 1.5, 0]
scene.objects['pink_salon_chairs'].placements[2].rotation = [0, 0, 0]
scene.objects['pink_salon_chairs'].placements[3].position = [1.5, 1.5, 0]
scene.objects['pink_salon_chairs'].placements[3].rotation = [0, 0, 0]

# Add round mirrors on the walls above the chairs
round_mirrors = add_object("large round mirror", num=4)
scene.objects['round_mirrors'] = round_mirrors

scene.objects['round_mirrors'].placements[0].position = [-1.5, 1.5, 1.5]  # Positional match to the chair
scene.objects['round_mirrors'].placements[0].rotation = [0, 0, 0]
scene.objects['round_mirrors'].placements[1].position = [-0.5, 1.5, 1.5]
scene.objects['round_mirrors'].placements[1].rotation = [0, 0, 0]
scene.objects['round_mirrors'].placements[2].position = [0.5, 1.5, 1.5]
scene.objects['round_mirrors'].placements[2].rotation = [0, 0, 0]
scene.objects['round_mirrors'].placements[3].position = [1.5, 1.5, 1.5]
scene.objects['round_mirrors'].placements[3].rotation = [0, 0, 0]

# Add pot plants
pot_plants = add_object("stylish pot plant", num=4)
scene.objects['pot_plants'] = pot_plants
scene.objects['pot_plants'].placements[0].position = [-2.9, 3.9, 0]  # Corners of the room
scene.objects['pot_plants'].placements[0].rotation = [0, 0, 0]
scene.objects['pot_plants'].placements[1].position = [-0.9, 3.9, 0]
scene.objects['pot_plants'].placements[1].rotation = [0, 0, 0]
scene.objects['pot_plants'].placements[2].position = [1.1, 3.9, 0]
scene.objects['pot_plants'].placements[2].rotation = [0, 0, 0]
scene.objects['pot_plants'].placements[3].position = [2.9, 3.9, 0]
scene.objects['pot_plants'].placements[3].rotation = [0, 0, 0]

# Update materials
scene.objects['walls_0'].material = retrieve_material("smooth light pink plaster wall")
scene.objects['floors'].material = retrieve_material("glossy light wood flooring")
scene.objects['ceilings'].material = retrieve_material("sleek white matte ceiling")

# Add modern lighting
modern_ceiling_lights = add_object("modern ceiling light fixture", num=1)
scene.objects['ceiling_lights'] = modern_ceiling_lights
scene.objects['ceiling_lights'].placements[0].position = [0, 0, 3.5]  # Center of the ceiling
scene.objects['ceiling_lights'].placements[0].rotation = [0, 0, 0]
# Define the ceiling light parameters
modern_lighting = Lighting(
    is_ceiling_light=True,
    light_type="point",
    energy=700,  # Bright energy
    color=[1.0, 1.0, 1.0]  # Neutral white light for natural illumination
)

# Attach the lighting to the ceiling light fixture
scene.objects['ceiling_lights'].lighting = modern_lighting
\end{lstlisting}

\subsection{\modulec}
We provide the base prompts we fed to our VLM policy (i.e., Molmo-7B~\cite{deitke2024molmo}) at each step of the iteration:

- Description prompt
\begin{lstlisting}[]
You are iteratively placing small objects on the {base_object_description} through multiple rounds
Your current goal is to add one object to the current scene as shown in the image.
Give a concrete description in less than 15 words of the appearance of a small object that makes sense to be placed on {base_object_description} given what you already see placed.
No need to justify your choice, just describe the placed object in less than 15 words.
\end{lstlisting}

- Location prompt
\begin{lstlisting}[]
You are looking at the following object: {base_object_description}.
Point to one location where it makes the most sense to place {retrieved_object_description} such that it does not collide with other objects but could even be on inside or on top of another object.
\end{lstlisting}

\begin{figure}[h]
    \centering
    \includegraphics[width=.7\linewidth]{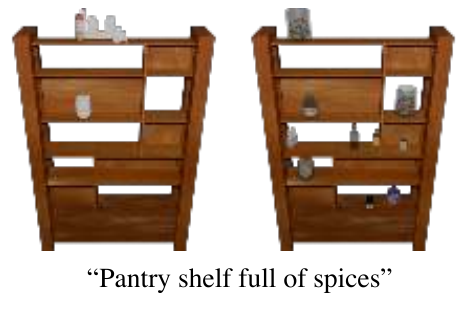}
    \caption{We experimented with two approaches for the asset level policy: prompt the VLM to point to a location then prompt for a description of the object to add (left) or prompt the VLM for a description of the next object to add then prompt it for a location (right). Note that in the right image, we provide the object as context to the VLM which helps to better space out the placements.}
    \label{fig:asset_level_ablation}
\end{figure}

Initially, we prompted the VLM for \(\mathcal{O}\), a set of plausible placement locations, and then iterated through them, prompting the VLM for descriptions of objects to place at each location \(o'_i\). This did not work as the pointing mechanism of the VLM tended to cluster points nearest the camera, and much of the object was occluded from view.
Empirically, randomizing view points can alleviate this issue. We also found qualitatively that it is important to use the VLM prior to decide where to place an asset given the language description of the asset, as illustrated in Figure~\ref{fig:asset_level_ablation}.

Our final implementation prompts the VLM for an object to place then prompts for a pixel location.



\section{Details of our Experiments}
We make design choices for our experiments to remain within a reasonable time and compute budget. Instead of leveraging text or image-to-3D asset generators~\cite{zhang2024clay}, diffusion models for PBR material synthesis, or the latest state-of-the-art Vision-Language Models to enhance generation quality, we rely on asset and material repositories for retrieval and use GPT-4o as the base model (except that Molmo-7B~\cite{deitke2024molmo} is used to output pixel locations in \textit{\modulec}) and its fine-tuning API.
In all experiments and baselines, we applied a combination of manual and automated filtering techniques to Objaverse~\cite{deitke2023objaverse}, resulting in a curated set of 200,000 3D assets, following the methodology outlined in~\cite{tang2024lgm}. Additionally, we scraped ambientCG\footnote{\url{https://ambientcg.com/}}, obtaining 2,000 PBR materials. It is important to note that all baselines utilized the same asset and material repositories.

Below, we first detail the two chosen general image-domain benchmarks commonly used to evaluate Vision-Language Models in one of our experiments in evaluating \method's \textit{\moduleb}. Subsequently, we discuss our implementation of the baseline in our experiment for \textit{\modulec}.

\subsection{VLM Benchmarks}
Here we detail the two benchmarks we use in section 4.1 to evaluate \method's \textit{\moduleb}'s visual grounding abilities.
\begin{description}
    \item[] \textbf{Object HalBench}~\cite{rohrbach2018object} compares the objects mentioned in the model's generations to the annotated objects from COCO images~\cite{coco}, thus providing an object-level hallucination rate for the model's image descriptions. We follow \cite{yu2024rlhf}'s implementation with augmented prompting and \texttt{gpt-4-turbo} judging hallucinations, and report the CHAIR scores (frequency of hallucinatory objects in model responses) at both the response level (CHAIRs) and object level (CHAIRi).
    \item[] \textbf{AMBER}~\cite{wang2023llm} assesses the VLM's hallucinations with fine-grained object, attribute, and relation annotations. We follow the official implementation for the generative task of AMBER and report the metrics CHAIR, COVER (ground-truth object coverage), HAL (rate of hallucinated responses), and COG (rate of hallucinatory objects similar to human cognition).
\end{description}

\subsection{\moduleb's Evaluation}
\paragraph{Baseline choice} In our experiment, we compared to LayoutGPT \cite{feng2024layoutgpt} and Holodeck \cite{yang2024holodeck}. We were unable to compare against SceneCraft \cite{hu2024scenecraft}, since its source code is not available. Although AnyHome \cite{fu2024anyhome} has released part of the source code, the full release has not yet occurred. PhyScene \cite{yang2024physcene} is not an \textit{open-vocabulary} layout generation method, it can only generate layout for a fixed set of categories.

\subsection{\modulec's Evaluation}
ProcTHOR~\cite{deitke2022} also has a method to place small objects on surfaces of ``receptacle objects". However, rather than being placed according to a natural language prompt, objects are placed based on the frequency that a given object type appears on the receptacle in the hand-modeled AI2thor and RoboTHOR datasets. Thus, we opt to compare with an inpainting-based baseline mainly.

As ARCHITECT~\cite{wang2024architectgeneratingvividinteractive}'s code is unreleased at the time of writing this paper, we implemented a similar method that inpaints the placeable areas, uses a VLM to identify the objects, gets each object's bounding box to determine placement location, then retrieves objects according to their description and places them. More specifically, the inpainting model takes as input a rendered view of the object and a mask of the areas to generate. We render a front or top-front view of the ``receptacle asset" and generate a mask of the placeable areas (on a table, between shelves) by placing a cube in or on top of the asset slightly smaller than the object's bounding box. GPT-4o is then used to create a list of objects to prompt GroundingDino~\cite{ren2024grounded} with the inpainted image to retrieve the bounding boxes. Utilizing the pixel location of the center of the bottom of the bounding box and the camera parameters, we generate a 3D ray to get a precise placement location. To remain consistent with our iterative method, we restrict placement to a maximum of 10 of the identified objects.



\subsection{GPU cost}
Unlike existing monolithic pipelines, \method's iterative design allows rendering and reasoning over complex scenes in a way that can scale with compute. All our experiments are conducted with A100s, with 1-3 minutes needed to generate the 3D scenes shown in Figure 1. 




